\documentstyle[aps,prl,preprint,amssymb,epsfig,floats,times]{revtex}
\begin{document}
\draft
\tightenlines
\title{Pattern formation in inclined layer convection}

\author{Karen E. Daniels, 
Brendan B. Plapp,\cite{bbp:address}
and Eberhard Bodenschatz\cite{eb:email}}

\address{Laboratory of Atomic and Solid State Physics, Cornell
University, Ithaca, NY 14853-2501}

\date{\today}
\maketitle

\begin{abstract}
We report experiments on thermally driven convection in an inclined
layer of large aspect ratio in a fluid of Prandtl number $\sigma
\approx 1$. We observed a number of new nonlinear, mostly 
spatio-temporally chaotic, states. At small angles of inclination 
we found longitudinal rolls, subharmonic oscillations,
Busse oscillations, undulation chaos, and crawling rolls. At 
larger angles, in the vicinity of the transition from buoyancy- 
to shear-driven instability, we observed drifting transverse rolls, 
localized bursts, and drifting bimodals. For angles past vertical, 
when heated from above, we found drifting transverse rolls and 
switching diamond panes. 
\end{abstract}


Rayleigh-B\'enard convection (RBC) of a vertical fluid layer heated
from below has long served as a paradigm for pattern forming systems
\cite{Cross93:pf,Boden99:review}. Variations that alter the
symmetries, such as rotation around a vertical axis 
\cite{Bajaj:1998:SPR,Hu:1998:CUR} and vertical vibrations 
\cite{Rogers99:vertosc_conv} continue to lead to important
insights. Another case, of particular meteorological and oceanographic
interest, is RBC of a fluid layer inclined with respect to gravity. 
This system is not only well suited for the study of buoyancy and
shear flow driven instabilities, but may also serve, along with liquid
crystal convection \cite{Kramer:1995:CIN}, as a paradigm for
anisotropic pattern forming systems.

\begin{figure}
\centerline{\epsfig{file={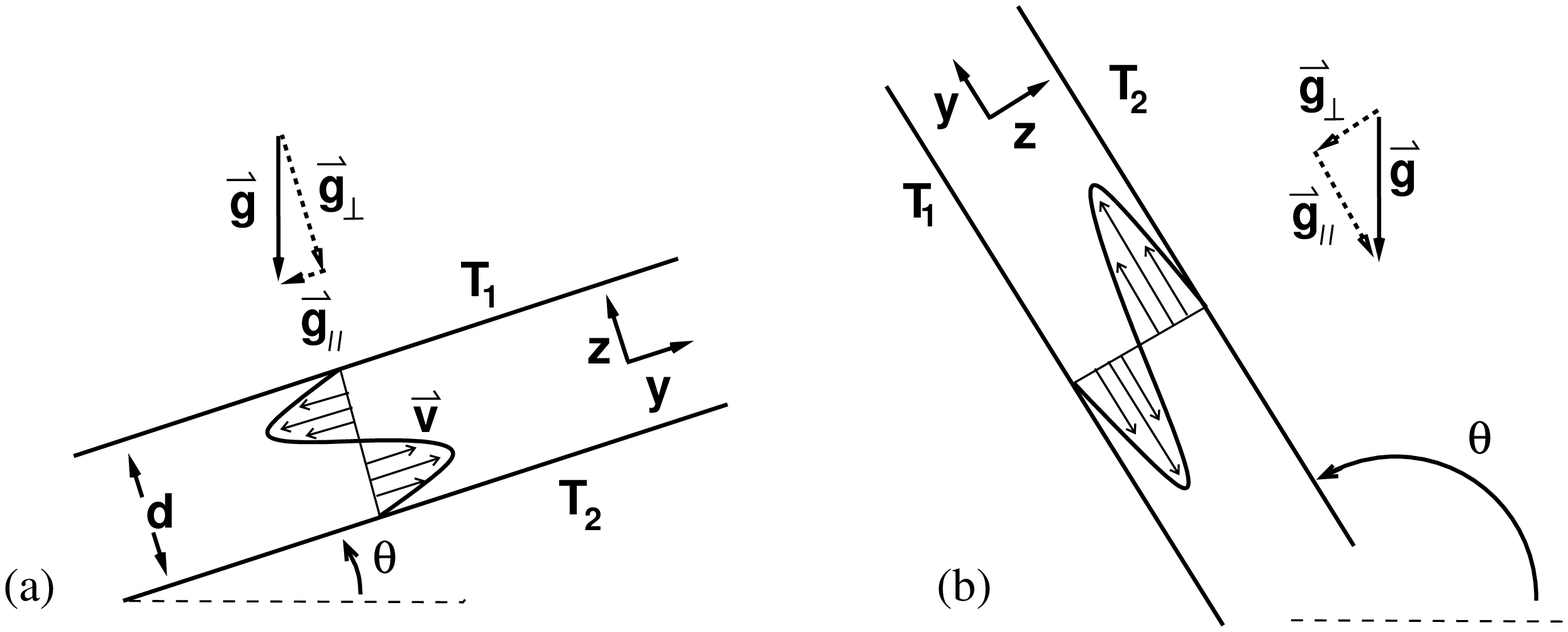}, width=5.5in}}
\bigskip
\caption{Schematic of the base flow. (a) Heated from below and (b) heated 
from above; cell height $d$, gravitational acceleration ${\bf g}$, 
shear flow ${\bf v}$, and temperature difference 
$\Delta T \equiv T_2 - T_1$ with  $T_2 > T_1$.} 
\label{f:profile}
\end{figure}

As with RBC, the onset of inclined layer convection (ILC) occurs when
the temperature difference, $\Delta T$, 
across the layer is sufficient for convection rolls to form.
The main difference from RBC is that, in ILC, the
patternless base state is characterized not only by a linear
temperature gradient but also by a symmetry-breaking shear flow. As
shown in Fig.~\ref{f:profile}, the component of gravity tangential to
the fluid layer, ${\bf g}_\parallel$, 
causes buoyant fluid to flow up along the warm plate
and down along the cold plate. For small angles of inclination
$\theta$, buoyancy dominates over shear flow (${\bf g}_\perp$ large, 
${\bf g}_\parallel$ small), and the primary instability is to {\it
longitudinal rolls} (LR) whose axes are aligned with the shear flow
direction \cite{Chen89:slot}. With increasing $\theta$, buoyancy effects 
decrease and for
$\theta > 90^\circ$ buoyancy is stabilizing. Above a critical angle
$\theta_{\mathrm c}$ ($\lesssim 90^\circ$) the shear flow causes a 
primary instability to {\it transverse rolls}
(TR) with roll-axes perpendicular to the shear flow 
\cite{Chen89:slot}. The few prior experiments
\cite{Hart71:incline,Ruth80:incline,Shadid90:incline} on ILC showed
reasonable agreement with the linear theory
\cite{Chen89:slot,Fujimura93:both,Clever77:incline,Busse92:incline}.
These experiments also demonstrated that LR are unstable to
some form of undulations
\cite{Hart71:incline,Ruth80:incline,Shadid90:incline}, in qualitative
agreement with theory \cite{Clever77:incline,Busse92:incline}, but the 
quantitative details of the state were
inaccessible due to experimental limitations. 

Here we report the first experimental results on pattern formation in
ILC for large aspect-ratio systems  in a range of inclination angles 
$0^\circ \leq \theta \leq 120^\circ$, {\it i.e.}, 
from horizontal (heated from below) to
past vertical (heated from above). We found many unpredicted
states when increasing $\Delta T$ above the critical temperature
difference.
For $0^\circ \leq \theta \lesssim 77.5^\circ$ we observed 
longitudinal rolls, subharmonic oscillations, Busse oscillations,
undulation chaos, and crawling rolls. In the neighborhood of the
codimension two point \cite{Fujimura93:both} for thermal and shear driven 
instability ($77.5^\circ \lesssim \theta \lesssim 84^\circ$), we
observed drifting bimodals, drifting transverse rolls, and localized
longitudinal and transverse bursts. For inclinations $\theta \gtrsim
84^\circ$ we found drifting transverse rolls, switching diamond-panes,
and longitudinal bursts. Most of these novel states were
spatio-temporally chaotic and were found very close to onset, where
theoretical progress should be possible.


{\bf Experiment:} Our experimental apparatus consisted of a
water-cooled pressure chamber containing a convection cell of diameter
10 cm, subdivided into two large aspect ratio rectangular cells. The
experimental design was similar to the one described in
 \cite{deBruyn96:apparatus}. The optically flat upper and lower
plate of the convection cell consisted of $1$ cm thick single
crystal sapphire and single crystal silicon, respectively. The
sapphire plate was cooled by a water bath, while the silicon plate was
heated by an electric film heater. The convection patterns were
visualized by the usual shadowgraph technique
\cite{deBruyn96:apparatus}. The
sidewalls were constructed of 9 layers of notebook paper, providing
the best possible
thermal matching between cell boundaries and the fluid. As measured 
interferometrically, the plates were parallel to $\pm 0.5$ $\mu$m. The
convection cell was housed in a pressure chamber, which held both the
cooling water and the convecting gas to ($41.37 \pm 0.01$) bar,
regulated to $\pm 5 \times 10^{-3}$ bar. The temperatures of the two
plates were regulated to $\pm 0.0003^\circ$C. 
Throughout the experiment the mean temperature was
kept constant at $(27.00\pm0.05)^\circ$C. We determined the cell height
$d$ by measuring the pattern wavenumber at onset for $\theta <
60^\circ$ and comparing it with the theoretical value of $q_c = 3.117
d$. We found $ d= (710 \pm 2) \mu$m and $d$ = $(702 \pm 2) \mu$m for
two sets of experiments.  The two convection cells had a size
$\Gamma_1 \simeq (21 \times 42) d^2$ and $\Gamma_2 \simeq 
(14 \times 48) d^2$.  For all data, the Prandtl number was 
$\sigma \equiv \nu/\kappa = 1.07$ as determined from 
\cite{deBruyn96:apparatus}, with the kinematic 
viscosity $\nu$ and thermal diffusivity
$\kappa$. The vertical thermal diffusion
time was $\tau_{\mathrm v} \equiv d^2/\kappa = 3.0$ s. Inclines from
$0^\circ$ (horizontal) to $120^\circ$ ($30^\circ$ past vertical) were
possible, with an accuracy of $\pm 0.02^\circ$. 
Following \cite{Boden99:review} we calculated the Boussinesq number 
${\cal P}(\theta)$ for the corresponding horizontal layer to estimate 
non-Boussinesq effects. At $\Delta T_{\mathrm c}(\theta)$ for 
$\theta < 70^\circ$ we found ${\cal P}(\theta) < 1.0$, putting the
flow into the Boussinesq regime. For larger angles ${\cal P}$ increased
linearly to 3.0 for the largest temperature differences investigated.
We observed the same convective patterns in both convection cells.


{\bf Onset of convection:} In ILC, the forward bifurcation to
LR is predicted to occur at the critical Rayleigh
number $R_{\mathrm c}(\theta) = R_{\mathrm c}^t(0^\circ) / {\cos
\theta}$ where $R_{\mathrm c}^t(0^\circ) = 1708 = \alpha d^3 g \Delta
T_{\mathrm c} / \kappa\nu$ ($\alpha$ is the thermal expansion
coefficient and $\Delta T_{\mathrm c}$ is the critical temperature 
difference). The threshold for the forward bifurcation to shear driven 
TR at large inclination angle is more 
complicated, and can only be determined
numerically \cite{Clever77:incline,Busse92:incline,Pesch:pers}.

\begin{figure}
\centerline{\epsfig{file={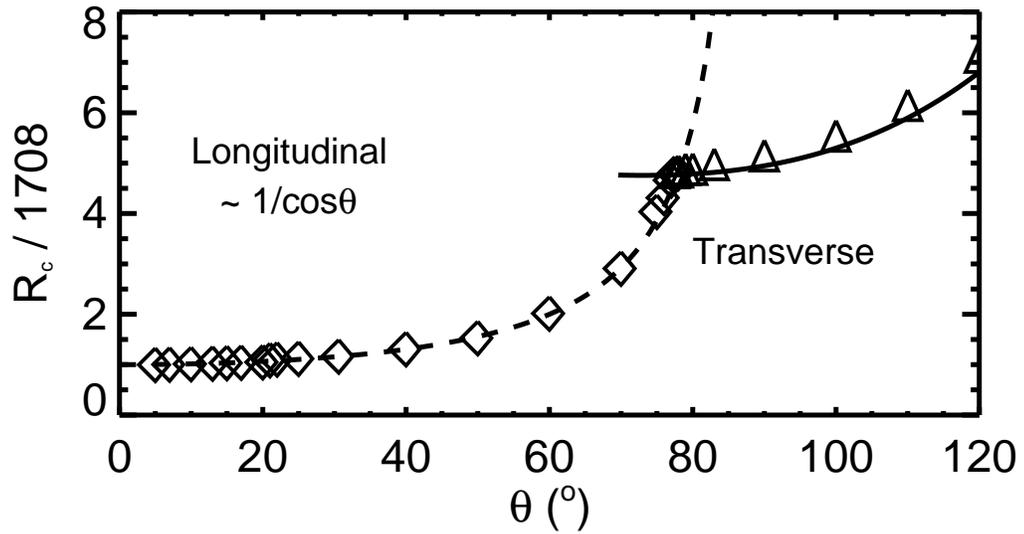}, width=5.5in}}
\bigskip
\caption{Onset of longitudinal rolls ($\diamond$) and drifting  
transverse rolls ($\triangle$). Also plotted are the predicted onsets 
for longitudinal (dashed) and transverse rolls (solid)
\protect\cite{Pesch:pers}.} 
\label{f:onsets}
\end{figure}
\begin{figure}
\centerline{\epsfig{file={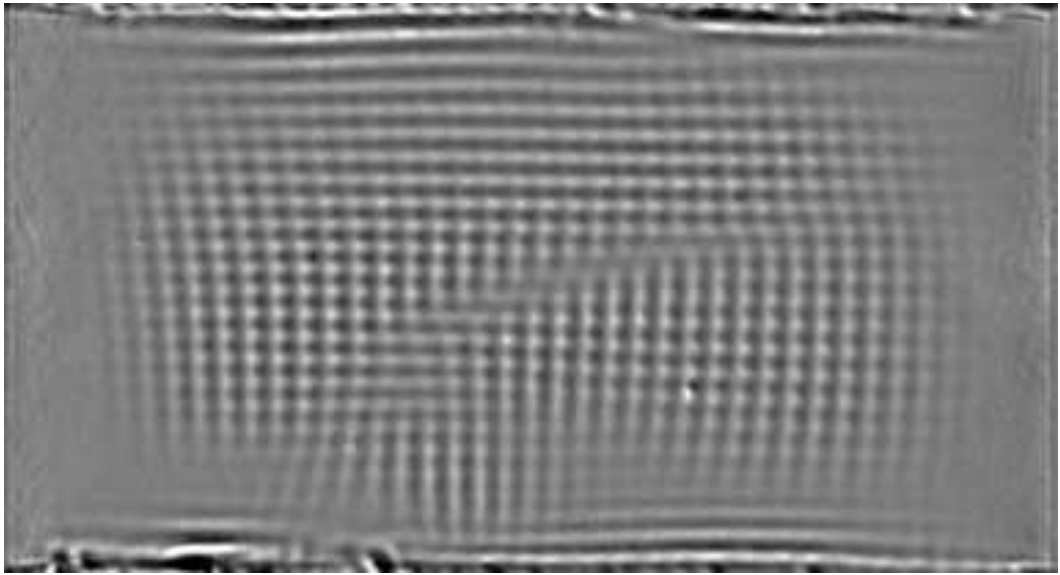}, width=5.5in}}
\bigskip
\caption{Digitally enhanced shadowgraph image of bimodals 
drifting from left to right in
Cell 1 for $\theta=77.6^\circ$, $\epsilon=0.01$. The left side of the
cell is higher than the right, with warm up-flow to the left and cold
down-flow to the right. The rolls at the edges of the cell are caused 
by sidewall imperfections. MPEG movies of all states are available
online \protect\cite{website}.}
\label{f:squares}
\end{figure}

We determined $\Delta T_{\mathrm c}$ for convection by
quasi-statically increasing $\Delta T$ in steps of 1 mK every 
$20$ minutes past the point where convection was observable and then 
decreasing the temperature difference similarly.  For all angles we 
observed forward bifurcations. Figure \ref{f:onsets} shows the 
measured $R_{\mathrm c}(\theta)$, as well as the theoretically 
predicted onsets for both the
buoyancy-driven (longitudinal) and the shear-driven (transverse)
instabilities \cite{Pesch:pers}. We found agreement with
theory for the onset of LR: the experimentally observed value was
$R_{\mathrm c}(\theta) = R_{\mathrm c}^e(0^\circ) / {\cos \theta}$
with $R_{\mathrm c}^e(0^\circ) = 1687 \pm 24$. We did not, however,
observe the theoretically predicted stationary TR, but instead 
drifting TR (DTR) at a slightly larger critical Rayleigh number. The
drift down the incline may be attributed to the broken symmetry across
the layer which is caused by the temperature dependence of the fluid
parameters (non-Boussinesq effects). Very interesting is the vicinity
of the theoretically predicted codimension two point at
$\theta_{\mathrm c} = 77.76^\circ$\cite{Pesch:pers}, where LR and TR 
have the same onset value.  Experimentally, we found a
forward bifurcation to DTR above $\theta_{\mathrm c} = (77.5 \pm
0.05)^\circ$, and in the range $77.5^\circ \leq \theta \leq 84^\circ$
DTR lost stability to {\it drifting bimodals} (DB)  above 
$\epsilon \approx 0.001$.  As shown in
Fig.~\ref{f:squares}, DB consist of a superposition of LR and DTR. Here
$\epsilon \equiv {{(\Delta T - \Delta T_{\mathrm c}(\theta))} /
\Delta T_{\mathrm c}(\theta)}$ is the reduced control parameter.
Theoretically, Fujimura and Kelly \cite{Fujimura93:both} predicted a
forward bifurcation to transverse rolls, which lose stability to
bimodals at $\epsilon \approx 0.001$ in a narrow angular region. We
find good agreement with these predictions, but with the difference that
the experimentally observed patterns are drifting.

\begin{figure}
\centerline{\epsfig{file={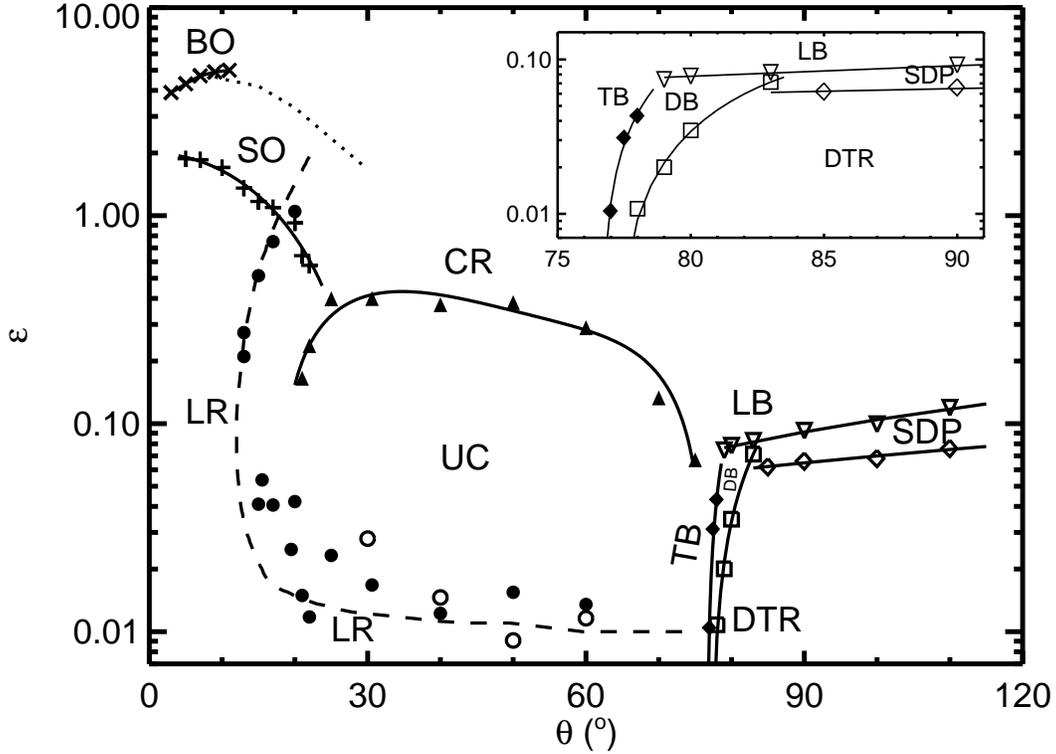}, width=5.5in}}
\bigskip
\caption{($\epsilon$, $\theta$) phase-space showing the boundaries
between the different nonlinear states. LR (longitudinal rolls), BO
(Busse oscillations), SO (subharmonic oscillations), UC (undulation
chaos), CR (crawling rolls), DTR (drifting transverse rolls), DB
(drifting bimodals), LB (longitudinal bursts), TB (transverse bursts),
and SDP (switching diamond panes). The dotted line is the predicted
onset of Busse oscillations for $\sigma=0.7$
\protect\cite{Clever77:incline}, the dashed line is the predicted
onset of undulations \protect\cite{Pesch:pers}, and the solid lines
are guides to the eye. Open circles (UC) were measured via defect
density \protect\cite{KED99:prep}, open diamonds (SDP) 
were measured via correlation
length \protect\cite{KED99:prep}, and the remainder were measured
visually.  The inset
shows a magnification of the codimension-two region.} 
\label{f:phaseplot}
\end{figure}
\begin{figure}
\centerline{\epsfig{file={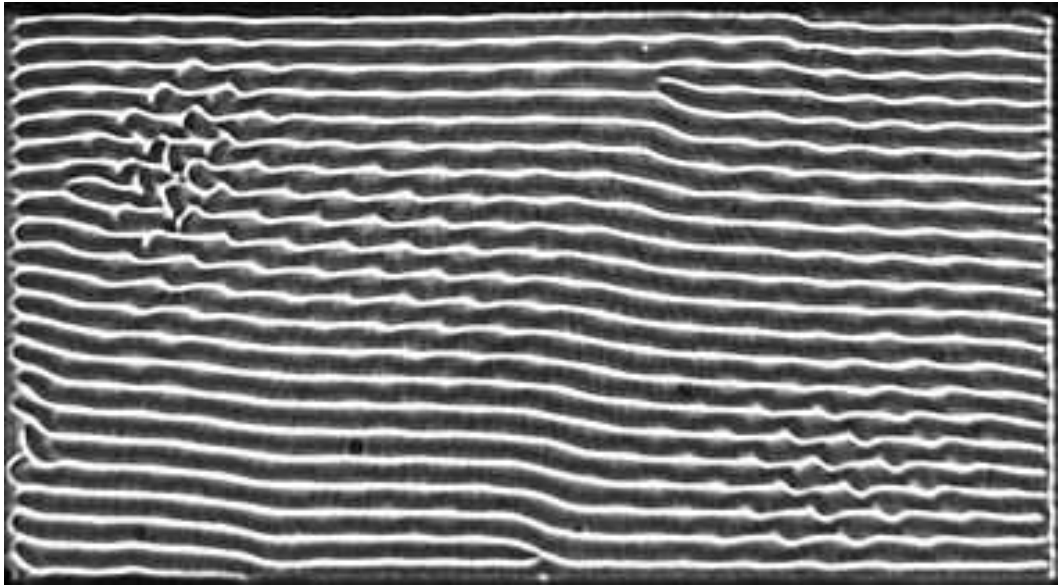}, width=5.5in}}
\bigskip
\caption{Contrast enhanced shadowgraph image of subharmonic
oscillations in Cell 1, with a turbulent burst in the upper left
corner. $\theta=17^\circ$, $\epsilon=1.5$. Warm up-flow to 
the left and cold down-flow to the right.}
\label{f:subosc}
\end{figure}

{\it \bf{Nonlinear states:}} Figure \ref{f:phaseplot} shows the measured
phase boundaries for the ten observed nonlinear convective states. 
At low angles ($\theta < 13^\circ$), LR are
stable up to $\epsilon \simeq 1$ above which the novel state of {\it
subharmonic oscillations} (SO) sets in. These oscillations are
characterized by a pearl-necklace-like pattern of bright (cold) spots
that travel along a standing wave pattern of wavy rolls. As shown in
Fig.~\ref{f:subosc}, these oscillations appear in patches whose location 
changes in time. Typical frequencies of the 
oscillations were measured to be 1 to 3 cycles per
$\tau_{\mathrm v}$. A recent theoretical analysis has shown agreement
with this value \cite{Busse:pers}. With further increase in
$\epsilon$, localized patches of traveling oscillations burst
intermittently. Within ${\cal O}(\tau_{\mathrm v})$ the amplitude of the
rolls' waviness increases, the pattern tears transverse to the rolls as
shown in the upper left corner of Fig.~\ref{f:subosc}, and fades away
leaving an almost parallel roll state.


For $\theta \simeq 10^\circ$ and $\epsilon \gtrsim 4$, we observed
patches of the well-known {\it Busse oscillations} (BO) coexisting with
patches of the SO. As shown by the dotted line in
Fig.~\ref{f:phaseplot}, our data for the onset of the
BO agrees well with the theoretical prediction calculated for $\sigma=0.7$
\cite{Clever77:incline}. 
It is surprising, however, that both oscillations (SO and BO)
coexist as localized patches in the same cell.

\begin{figure}
\centerline{\epsfig{file={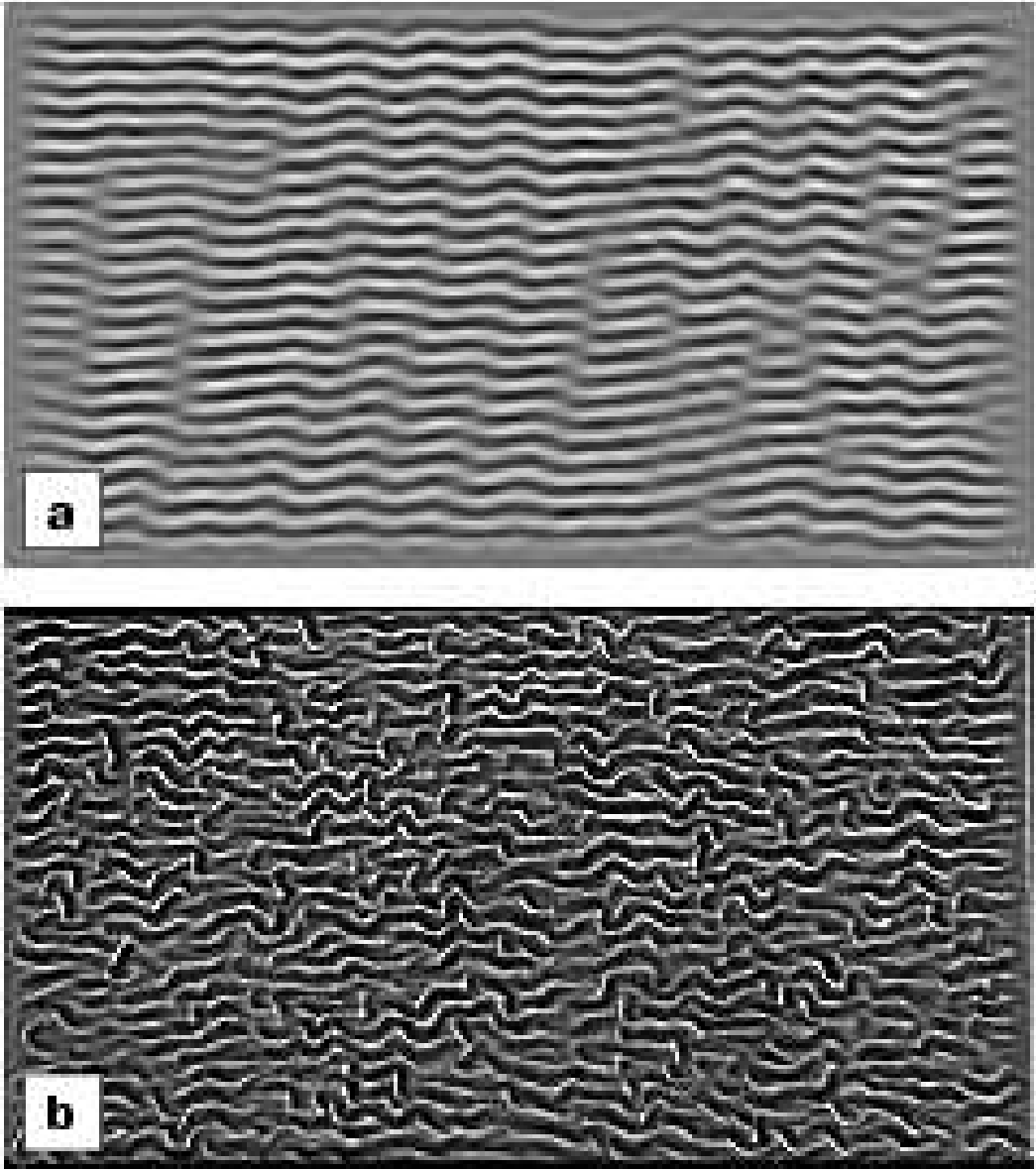}, width=5.5in}}
\bigskip
\caption{Digitally enhanced shadowgraph images of convection
states at $\theta = 40^\circ$ in Cell 1. (a) Undulation chaos at
$\epsilon=0.07$. (b) Crawling rolls at $\epsilon=0.88$. Warm
up-flow to the left and cold down-flow to the right.} \label{f:midpics}
\end{figure}

At intermediate angles ($25^\circ < \theta < 70^\circ$), where the
initial instability is to LR we found with increasing $\epsilon$ that
LR were unstable to undulations. Although the
experimentally determined value for the instability $\epsilon \approx
0.01$ agrees well with the theoretical prediction 
(see Fig.~\ref{f:phaseplot})
\cite{Clever77:incline,Busse92:incline,Pesch:pers}, we did not observe
a stationary pattern of undulations, but a defect-turbulent
state of {\it undulation chaos} (UC). A snapshot of 
UC is shown in Fig.~\ref{f:midpics}a. 
At $\epsilon \gtrsim 0.11$, the UC begins to
``twitch'' in the direction transverse to the rolls on time scales
${\cal O}(\tau_{\mathrm v})$. With increasing $\epsilon$, the
amplitude of the twitching increases and the rolls eventually tear,
with the ends ``crawling'' in the direction transverse to the original
rolls. A snapshot of {\it crawling rolls} (CR) is shown in
Fig.~\ref{f:midpics}b.

\begin{figure}
\centerline{\epsfig{file={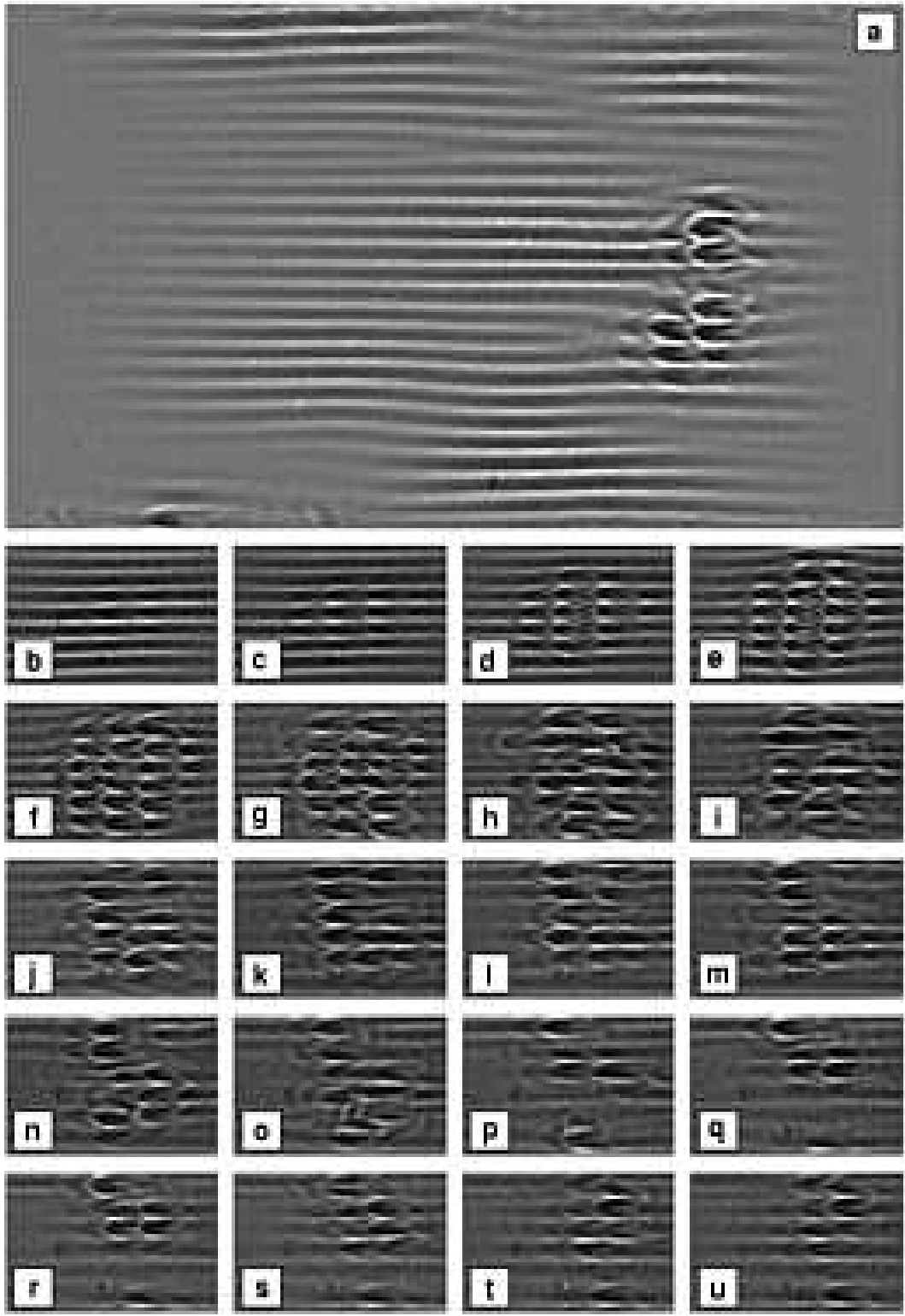}, width=5.5in}}
\bigskip
\caption{Digitally enhanced images of (a) transverse bursts in
spatio-temporally chaotic longitudinal rolls, at $\theta=77^\circ$ and 
$\epsilon=0.04$ in Cell 1. (b -- u) time-evolution of a single burst
at time-intervals $0.18\tau_{\mathrm v}$. Warm up-flow to 
the left and cold down-flow to the right.}
\label{f:transburst}
\end{figure}
\begin{figure}
\centerline{\epsfig{file={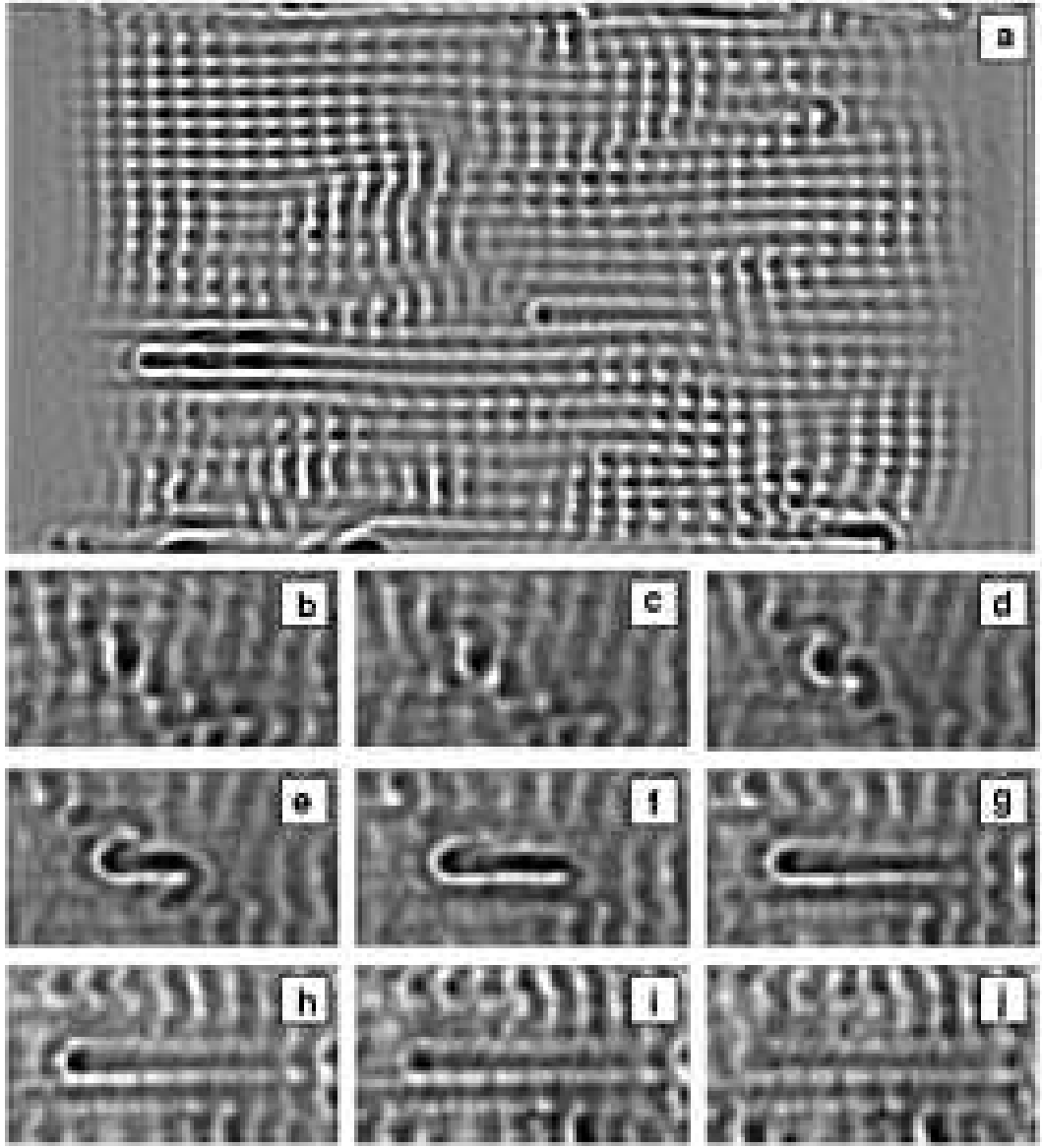}, width=5.5in}}
\bigskip
\caption{Digitally enhanced images of (a) longitudinal bursts 
at $\theta=79^\circ$ and 
$\epsilon=0.10$ in Cell 1. (b -- j) time-evolution of a single burst
at time-intervals $0.09\tau_{\mathrm v}$. Warm up-flow to 
the left and cold down-flow to the right.}
\label{f:longburst}
\end{figure}

In the vicinity of the codimension-two point, at $\theta_{\mathrm c}$,
we observed drifting bimodals quite close to onset. As shown in
Fig.~\ref{f:phaseplot}, for small angles the existence region of the pure
DB is limited by localized {\it transverse bursts} (TB), while for
large angles by DTR. A snapshot of transverse bursts and the 
evolution of a single burst is shown in Fig.~\ref{f:transburst}. In
this region of phase space the LR occur in patches that grow and decay
intermittently while TB nucleate in high amplitude LR-regions. As shown in
the time series in Fig.~\ref{f:transburst}, TB grow over the period
of a few $\tau_{\mathrm v}$ and then decay rapidly. Above
$\epsilon \approx 0.8$ the DB are unstable to localized {\it
longitudinal bursts} (LB) as shown in Fig.~\ref{f:longburst}a. As
shown in Fig.~\ref{f:longburst}b--j, a few longitudinal rolls grow
locally to large amplitude and then quickly fade. With both types
of bursts, the bursts increase both in density and frequency when
$\epsilon$ is increased, eventually developing into a turbulent state
at $\epsilon \gtrsim 1$.

\begin{figure}
\centerline{\epsfig{file={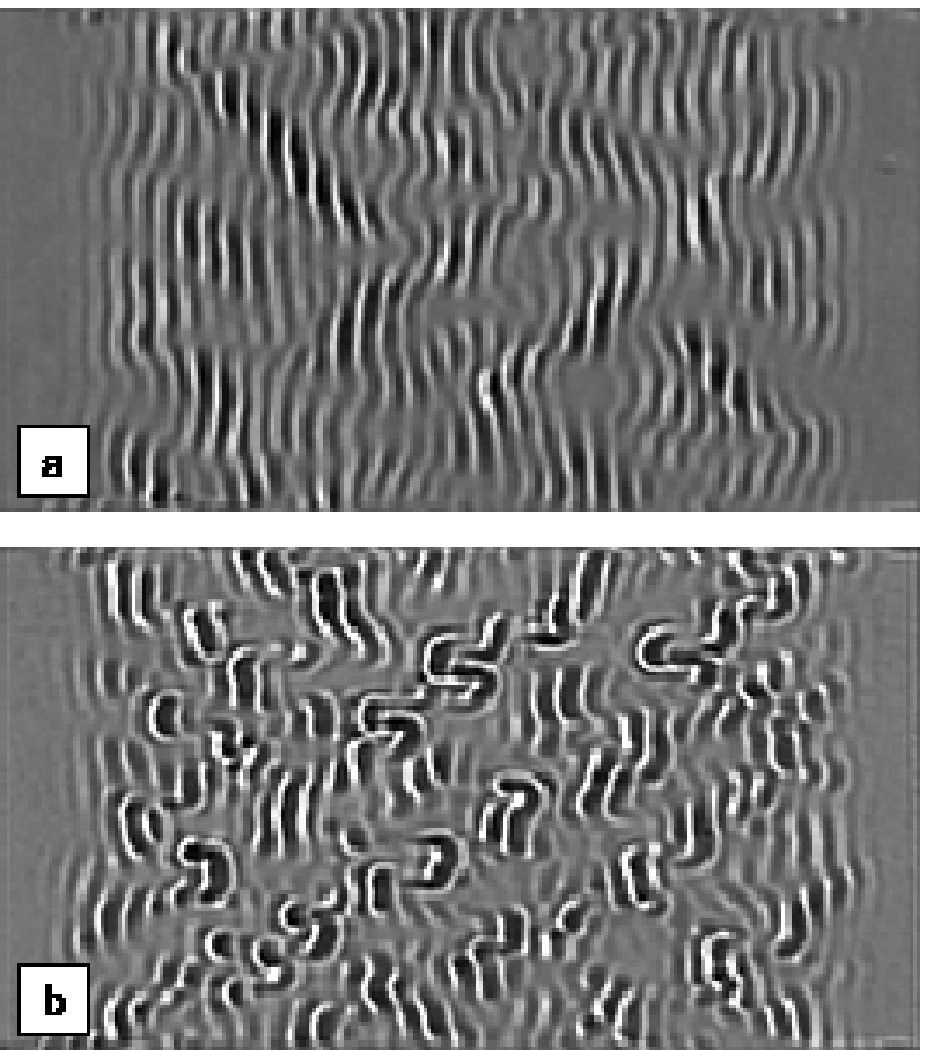}, width=5.5in}}
\bigskip
\caption{Digitally enhanced shadowgraph images of (a) 
switching diamond panes ($\epsilon=0.1$, 
$\theta=100^\circ$) and (b) longitudinal bursts within diamond panes
($\epsilon=0.19$, 
$\theta=100^\circ$) in Cell 1. Warm up-flow to 
the left and cold down-flow to the right.} 
\label{f:switching}
\end{figure}

Past $90^\circ$, we continued to observe shear driven convection
patterns. DTR are
the primary instability;  however, they are unstable to switching {\it
diamond panes} (SDP) at $\epsilon \approx 0.07$. The state is
characterized by spatio-temporally chaotic switching on time-scales of
${\cal O}(\tau_{\mathrm v})$ from $+45^\circ$ to $-45^\circ$ of large
amplitude regions of DTR, as seen in
Fig.~\ref{f:switching}a. At $\epsilon \gtrsim 0.1$ SDP are unstable to LB,
which in this region of phase space are denser but travel less
distance than in TR, as shown in Fig.~\ref{f:switching}b.


{\bf Conclusion:} Inclined layer convection in the weakly nonlinear
regime displays a rich phase diagram, with ten different states
accessible over the range of parameters investigated. The phase space
naturally divides into several regions of characteristic behavior
which have so far been characterized semi-quantitatively. All states
but longitudinal and transverse rolls are spatio-temporally
chaotic. Most instabilities occurred very close to onset and further
theoretical description should be possible.
 Especially interesting is the bursting behavior,
which may be related to turbulent bursts in other shear flows
\cite{Knobloch:1999:BMH}.

We thank F. H. Busse and W. Pesch for important discussions on the
stability curves and theoretical descriptions of various states.
E.B. acknowledges the kind hospitality of Prof H. Levine at the 
University of California at San Diego where part of this manuscript 
was prepared. We gratefully acknowledge support from the NSF 
under grant DMR-9705410.

\newpage

\begin{thebibliography}{10}

\bibitem[*]{bbp:address}
Presently at Center for Nonlinear Dynamics, University of Texas at Austin

\bibitem[\dagger]{eb:email}
Email: eb22@cornell.edu

\bibitem{Cross93:pf}
M.~C. Cross and P.~C. Hohenberg, Rev. Mod. Phys. {\bf 65},  851
  (1993).

\bibitem{Boden99:review}
E. Bodenschatz, W. Pesch, and G. Ahlers, Annu. Rev. Fluid Mech. {\bf 32},
  709 (2000).

\bibitem{Bajaj:1998:SPR}
K.~M.~S. Bajaj, J. Liu, B. Naberhuis, and G. Ahlers, Phys. Rev. Lett. {\bf 81},
  806  (1998).

\bibitem{Hu:1998:CUR}
Y. Hu, W. Pesch, G. Ahlers, and R.~E. Ecke, Phys. Rev. E {\bf 58},  5821
  (1998).

\bibitem{Rogers99:vertosc_conv}
J.~L. Rogers, M.~F. Schatz, J.~L. Bougie, and J.~B. Swift, Phys. Rev.
  Lett. (to be published).

\bibitem{Kramer:1995:CIN}
L. Kramer and W. Pesch, Annu. Rev. Fluid Mech. {\bf 27},  515  (1995).

\bibitem{Chen89:slot}
Y. Chen and A.~J. Pearlstein, J. Fluid Mech. {\bf 198},  513
  (1989) and references therein.

\bibitem{Hart71:incline}
J.~E. Hart, J. Fluid Mech. {\bf 47},  547  (1971).

\bibitem{Ruth80:incline}
D.~W. Ruth, G.~D. Raithby, and K.~G.~T. Hollands, J. Fluid Mech.
  {\bf 96},  481  (1980).

\bibitem{Shadid90:incline}
J.~N. Shadid and R.~J. Goldstein, J. Fluid Mech. {\bf 215},  61
  (1990) and references therein.

\bibitem{Fujimura93:both}
K. Fujimura and R.~E. Kelly, J. Fluid Mech. {\bf 246}, 545
  (1993); in {\em Bifurcation Phenomena and Chaos in
  Thermal Convection}, edited by H.~H. Bau, L.~A. Bertram, and
  S.~A. Korpela, ASME: HTD, {\bf 214}, 73, (1992).

\bibitem{Clever77:incline}
R.~M. Clever and F.~H. Busse, J. Fluid Mech. {\bf 81},  107
  (1977).

\bibitem{Busse92:incline}
F.~H. Busse and R.~M. Clever, J. Eng. Math. {\bf 26},  1
  (1992).

\bibitem{deBruyn96:apparatus}
J.~R. de~Bruyn, {\it et al.}, Rev. Sci. Instrum. {\bf 67},  2043
  (1996).

\bibitem{Pesch:pers}
W. Pesch (private communication).

\bibitem{website}
MPEG movies of spatio-temporally chaotic states are available at
  http://milou.msc.cornell.edu/ILCmovies.

\bibitem{KED99:prep}
K.~E. Daniels and E. Bodenschatz (unpublished).

\bibitem{Busse:pers}
F.~H. Busse and R.~M. Clever (private communication).

\bibitem{Knobloch:1999:BMH}
E. Knobloch and J. Moehlis, in {\it Pattern formation in continuous and 
  coupled systems}, edited by M. Golubitzki, D. Luss, and
  S.~H. Strogatz (Springer-Verlag, Berlin, 1999).
   
\end{thebibliography}
\end{document}